\documentclass[conference]{IEEEtran}
\IEEEoverridecommandlockouts

\usepackage{cite}
\usepackage{amsmath,amssymb,amsfonts}
\usepackage{algorithmic}
\usepackage{graphicx}
\usepackage{float}
\usepackage{textcomp}
\usepackage{booktabs}
\usepackage{multirow}
\usepackage{subcaption}
\usepackage{hyperref}
\usepackage{microtype}
\usepackage{xcolor}
\usepackage{xurl}
\def\BibTeX{{\rm B\kern-.05em{\sc i\kern-.025em b}\kern-.08em
    T\kern-.1667em\lower.7ex\hbox{E}\kern-.125emX}}

\newcommand{\toolOrch}{Osmo\xspace}
\newcommand{\toolLogger}{OneLogger\xspace}
\newcommand{\toolRGS}{NVIDIA Mission Control Autonomous Job Recovery Service\xspace}

\usepackage{xspace}

\begin{document}

\title{Instant GPU Efficiency Visibility at Fleet Scale}

\author{
\IEEEauthorblockN{Connor Pedersen, Dong H. Ahn,
Michel Migdal, Collin Neale, Nik Konyuchenko}
\IEEEauthorblockA{NVIDIA\\
\texttt{\{connorp, donga, mmigdal, cneale, nkonyuchenko\}@nvidia.com}}
}

\maketitle
\begin{abstract}
We present Overall FLOP Utilization (OFU), a hardware-level, precision-agnostic GPU efficiency metric for AI workloads on HPC systems, derived from two on-chip performance counters: Tensor Pipe Activity and SM clock frequency. OFU requires no application instrumentation and works across GPU generations and numeric precisions. We characterize five properties of OFU approximation---tile quantization, floating-point precision scaling, clock sampling noise, Tensor Core clock domains, and non-tensor undercounting---through controlled GEMM experiments on H100 and GB200 across FP16, TF32, FP8, and NVFP4. After tile-quantization correction, OFU predicts application-level MFU to within $\leq$2 percentage points. Against 608 production training jobs, OFU achieves $r = 0.78$ correlation with application-level MFU and surfaces two framework-level FLOPs miscalculations. Deployed across large-scale GPU fleets, OFU has detected a $2.5\times$ efficiency regression and tracked precision-dependent utilization changes in mixed-precision pretraining. Our evaluation and operational experience suggest OFU is a practical, deployment-ready complement to application-level MFU for continuous fleet-wide efficiency monitoring.
\end{abstract}


\section{Introduction}

The current push toward one of the largest capital-expenditure cycles in
technology history has made GPU compute efficiency a critical factor in
economic viability. Microsoft spent \$64.6B in capex in FY2025,
Alphabet invested \$91.4B in 2025, and Meta spent
\$69.7B for the same period---with cumulative Big Tech Artificial Intelligence (AI) High Performance Computing (HPC) infrastructure investment on track to exceed \$2.8T through 2029
\cite{microsoft2025golden,alphabet2025q2,reuters2025metaCapex,reuters2025citiAIcapex}.
At this scale, each percentage point of GPU utilization recovered across
a large HPC fleet\footnote{We use \emph{fleet} to refer to the aggregate of all GPUs deployed across one or more HPC or data centers under a single administrative domain.} represents millions of dollars in value; conversely,
undetected inefficiencies can silently double the effective cost of
compute. Fleet-wide measurement of GPU utilization which is accurate, continuous,
and actionable is therefore essential to realizing the full potential of
AI HPC infrastructure investment.

Achieving this requires a utilization metric that can be deployed across
every workload on a massive, heterogeneous fleet---without instrumenting
applications, without modifying the software stack, with
well-characterized accuracy, and scalable across a wide range of GPU
generations---while remaining simple enough to integrate into
fleet-wide resilience and goodput services at multiple levels that detect
inefficiencies and drive optimization.
Existing measurement techniques, however, fall short of these demands.

In measuring GPU utilization, three broad classes of techniques exist.
First, user-level performance profiling tools such as NVIDIA Nsight
Compute, Nsight Systems, and open-source alternatives like PyTorch
Profiler \cite{pytorch_profiler}, DeepSpeed Flops Profiler
\cite{deepspeed_flops_profiler}, and HPCToolkit \cite{hpctoolkit}, measure
floating-point utilization with high fidelity. However, these are
per-workload tools: each user must instrument and profile their own job,
they impose non-trivial overhead on the profiled application, and they
are designed for targeted profiling runs, not continuous fleet-wide
monitoring. Second, framework-level algorithmic throughput estimates
represent an improvement: once implemented within a framework such as
Megatron-LM \cite{nvidia2025megatronlm}, PyTorch Lightning
\cite{pytorch_lightning}, or NeMo \cite{nvidia_nemo}, any workload built on that framework can report Model FLOPs Utilization (MFU) derived from
model-architecture FLOPs\footnote{Throughout this paper, we use FLOPs to denote a count of floating-point operations and FLOP/s to denote a rate (floating-point operations per second).} counts \cite{chowdhery2022palm}. However,
coverage remains fragmented across frameworks, and the FLOPs formulas
must be updated for each new training modality---dense,
mixture-of-experts, latent-space routing, multimodal pipelines---making
them brittle and error-prone as architectures evolve. Third, hardware
performance counters, exposed through DCGM
\cite{nvidia_dcgm_exporter}, offer a non-intrusive alternative
requiring no application instrumentation and no runtime overhead, but
lack a rigorous study of their accuracy and nuance, validation against
controlled benchmarks, and demonstrated scalability across GPU
generations.

In this paper, we present GPU utilization techniques with
well-characterized accuracy, designed for instant fleet-wide visibility
and downstream optimization, collectively called \emph{Overall FLOP
Utilization} (OFU). OFU comprises four complementary elements: a hardware-counter-based
metric that combines Tensor Pipe Activity with SM clock frequency to
instantly expose floating-point utilization across any NVIDIA GPU
generation, without application instrumentation or software-stack
modifications; error characterization against well-defined computational
kernels (GEMMs) across precisions and GPU architectures, establishing
bounded accuracy for the metric; a deep analysis of how these kernels
are mapped to hardware, including the nuances of over- and undercounting
of floating-point operations due to tile quantization and peak-TFLOP/s
specification discrepancies; and operational experience deploying these
techniques at fleet scale, demonstrating that they indeed unearth
actionable areas of inefficiency.

Specifically, this paper makes the following contributions:
\begin{itemize}
  \item A first-principles derivation of the OFU metric as a simple
    product of Tensor Pipe Activity and normalized SM clock frequency,
    yielding a precision-agnostic, architecture-agnostic proxy for MFU
    that requires no model-specific information.
  \item Correlation with application-level MFU across 608
    production training jobs on H100 GPUs, achieving $r = 0.78$.
  \item Error characterization through controlled GEMM experiments on
    H100 and GB200 GPUs across FP16, TF32, FP8, and NVFP4, quantifying
    all major sources of error in our approximation.
   \item Operational experiences of deploying OFU to a large GPU fleet including three case studies that led to spot and address inefficient GPU usage.   
\end{itemize}

Our evaluation shows that controlled GEMM experiments on H100 and GB200
across FP16, TF32, FP8, and NVFP4 predict application-level MFU to within
1--3~percentage points, tightening to $\leq$2~percentage points after
tile-quantization correction. Against 608 production training jobs on
H100 GPUs, OFU achieves $r = 0.78$ correlation with application-level
MFU and surfaces two distinct framework-level FLOPs miscalculations.
Operationally, OFU has been deployed across large-scale GPU fleets at
multiple integration levels---per-job dashboards, cluster-wide resilience
services, and automated goodput monitoring---where it detected a
$2.5\times$ efficiency regression in embodied-agent training and tracked
precision-dependent utilization changes across mixed-precision
foundation-model pretraining.

Overall, our techniques demonstrate that hardware performance counters,
when carefully validated and operationalized, provide a practical path
to instant, fleet-wide GPU utilization visibility---meeting the demands
that existing measurement approaches cannot.

\section{Many Techniques, No Fleet-Wide Solution}
\label{sec:motivation}

From the perspective of fleet infrastructure management---where the goal
is to maximize the return on GPU investment across all users and
workloads---we prepared a fleet-wide efficiency review for a collection
of large internal GPU clusters. The measured training MFU
averaged approximately 20\% over a two-week window, well below the
35--50\% range considered healthy. Understanding where the waste
originated required a utilization metric available across all workloads,
but from this infrastructure vantage point, we met with the following
limitations to answering this fundamental question.

Application-reported MFU requires explicit per-framework integration:
each training codebase must compute and emit FLOPs counts via a
reporting tool such as \toolLogger~\cite{nvidia_onelogger}. At the time
of our review, only approximately 20\% of fleet workloads had been
onboarded. The remaining 80\% of GPU-hours had no MFU measurement at
all, leaving the majority of the fleet invisible to efficiency
monitoring. Onboarding additional workloads requires
application-level code changes---a process that does not scale
across a diverse fleet with hundreds of users and configurations.

Even where application MFU was available, we found it could be
significantly incorrect. A 288-GPU DeepSeek-style~\cite{deepseek_v2} MoE training job
reported 54.27\% MFU---appearing to be one of the fleet's
best-performing workloads. When measured with hardware counters, however,
the job showed only 25.58\% MFU, a 112.2\% relative
discrepancy.
Investigation revealed that the framework's FLOPs counter did not
account for the model's latent-space projections, inflating the reported
FLOPs by ${\sim}3\times$. Such errors are increasingly common as
emerging and diverse model architectures---mixture-of-experts,
latent-space routing, multimodal pipelines---outpace the assumptions
baked into framework-level FLOPs formulas. Researchers understandably
prioritize model quality over throughput-calculation accuracy, and may
lack the domain expertise to maintain correct FLOPs accounting as
architectures evolve.

Given these limitations, the review team demanded a solution that
collects utilization without modifying training code or injecting
profiling hooks; deploys with minimal changes to the AI software stack,
preserving backward compatibility; covers all workloads---training and
inference---regardless of framework, model architecture, or GPU
generation; integrates readily into fleet-wide resilience and goodput services that
detect inefficiencies and drive optimization; provides bounded,
well-characterized accuracy against established benchmarks; and
identifies where its approximations fall short, charting a path toward
higher-fidelity measurement in future hardware and software stacks.

\section{Overall FLOP Utilization}
\label{sec:ofu}

The requirements from Section~\ref{sec:motivation}---no application
instrumentation, no software-stack modifications, coverage across all
workloads and GPU generations, bounded accuracy, and integration into
fleet-wide automation---demand a metric grounded entirely in signals the
hardware already exposes. Meeting these requirements simultaneously demands
three things: (a)~a universally available measurement substrate that
is independent of any training framework or model architecture, (b)~a
mapping from that substrate to floating-point throughput rooted in the
physical execution pipelines of the GPU, and (c)~a quantitative understanding
of where the resulting approximation is tight and where it breaks down.
We satisfy~(a) with on-chip performance counters available through
DCGM~\cite{nvidia_dcgm_exporter}, satisfy~(b) by deriving an MFU
estimator from the first-principles structure of GPU floating-point
pipelines, and satisfy~(c) through controlled kernel-level experiments
and production-scale validation. The
remainder of this section develops the core derivation: we first describe
the floating-point execution model of modern NVIDIA GPUs from first
principles, then show how a single hardware counter---Tensor Pipe
Activity---combined with SM clock frequency can yield a precision-agnostic
utilization metric we call \emph{Overall FLOP Utilization} (OFU).

\subsection{Floating-Point Execution in Modern GPUs}

A modern NVIDIA GPU organizes its compute resources into an array of
\emph{Streaming Multiprocessors} (SMs), each capable of issuing
instructions to several independent execution pipelines in
parallel~\cite{nvidia_volta_whitepaper}. Two pipeline families are
relevant to floating-point throughput: the \emph{CUDA-core} (FP/INT)
pipelines and the \emph{Tensor Core} pipelines.

Each SM contains scalar floating-point/integer (FP32, FP64, INT32)
functional units organized into processing blocks, each executing one
operation per thread per cycle across a warp of 32~threads. The number of
FP32 CUDA cores per SM has grown with each generation---64 on Volta
and Ampere (A100), and 128 on both Hopper and
Blackwell~\cite{nvidia_volta_whitepaper,nvidia_a100_whitepaper,nvidia_h100_wp,semianalysis_tensor_core_evolution_2025}.
These units handle general-purpose arithmetic: element-wise activations,
reductions, address computation, and any floating-point work that is not
matrix multiplication.

Beginning with Volta, NVIDIA introduced dedicated \emph{Tensor Cores}:
fixed-function matrix-multiply-accumulate (MMA) units that operate at the
warp level on small tiles to compute $D = A \times B + C$ in a single
instruction~\cite{nvidia_volta_whitepaper,semianalysis_tensor_core_evolution_2025}.
Because each MMA instruction retires an entire tile of output elements
rather than one element per thread, the per-cycle throughput is substantially higher than the scalar
pipeline for the same silicon area. On
the H100, for example, each SM delivers 4,096~FP16 FLOPs/cycle on the
Tensor Core pipeline versus 256~FP32 FLOPs/cycle on the CUDA-core
pipeline---a $16\times$ ratio even before accounting for the $2\times$
element width difference~\cite{nvidia_h100_wp}. This gap has widened with
every generation as NVIDIA has invested transistor budget predominantly in
Tensor Core throughput.

In any workload dominated by matrix multiplications---which describes
virtually all modern deep-learning training and inference---the vast
majority of achieved FLOPs flow through the Tensor Core pipeline. The
CUDA-core contribution to total FLOPs is negligible by comparison. This
asymmetry is the physical basis for our approach: monitoring Tensor Core
pipeline activity alone captures the dominant term in GPU floating-point
utilization, and any FLOPs missed by ignoring the scalar pipeline are
well within the noise floor of practical measurement.

\subsection{Tensor Core Evolution Across Architectures}
Tensor Cores first appeared in Volta (V100). At the warp level, a single HMMA instruction computes a $16\times16\times16$ mixed-precision MMA---multiplying two FP16 input tiles and accumulating into FP32---delivering 8,192 FLOPs per instruction \cite{nvidia_volta_whitepaper}. Subsequent architectures broadened support.

Ampere introduced BF16 and TF32 on the Tensor Cores \cite{nvidia_a100_whitepaper,nvidia_tf32_blog} and replaced HMMA with the \texttt{mma.sync} instruction, whose native warp-level tile is $16\times8\times16$ (FP16/BF16 inputs, FP32 accumulation). Ampere also added \texttt{cp.async} for asynchronous global-to-shared-memory loads, enabling multi-stage pipelining crucial to saturating Tensor Cores.

Ada and Hopper added native FP8 (E4M3/E5M2). Hopper introduced asynchronous WGMMA instructions that increase parallelism and efficiency by accepting inputs in shared memory and supporting even larger tile sizes \cite{nvidia_hopper_wp,nvidia_hopper_fp8_blog,hopper_wgmma_paper}. Most recently, Blackwell added native FP4 via NVFP4 and tensor memory with tcgen05 instructions \cite{nvidia_nvfp4_blog,nvidia_blackwell_arch,nvidia_nvfp4_pretraining}.

A key trend across all generations is the widening gap between Tensor
Core and CUDA-core throughput: dense FP16 Tensor Core throughput has
grown from 125~TFLOP/s on V100~\cite{nvidia_volta_whitepaper} to
989~TFLOP/s on H100~\cite{nvidia_h100_wp}, while FP32 CUDA-core
throughput has increased far more modestly. This gap makes Tensor Core
activity the dominant signal for GPU compute utilization in modern AI applications.

\begin{figure*}[!t]
  \centering
  \begin{subfigure}[t]{0.24\textwidth}
    \includegraphics[width=\linewidth]{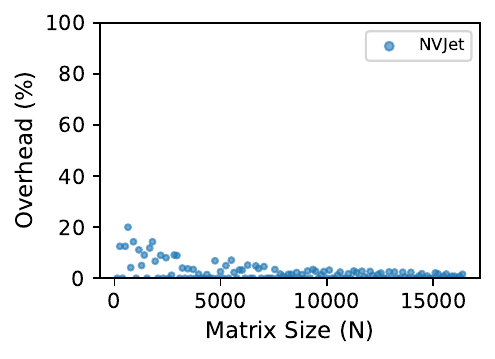}
    \caption{H100 FP16}
  \end{subfigure}\hfill
  \begin{subfigure}[t]{0.24\textwidth}
    \includegraphics[width=\linewidth]{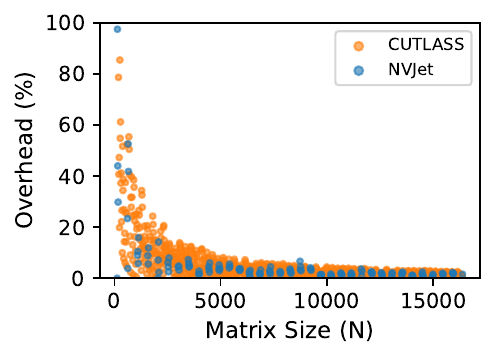}
    \caption{H100 FP16 (random)}
  \end{subfigure}\hfill
  \begin{subfigure}[t]{0.24\textwidth}
    \includegraphics[width=\linewidth]{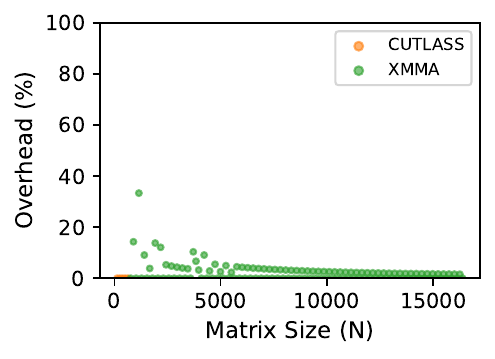}
    \caption{H100 TF32}
  \end{subfigure}\hfill
  \begin{subfigure}[t]{0.24\textwidth}
    \includegraphics[width=\linewidth]{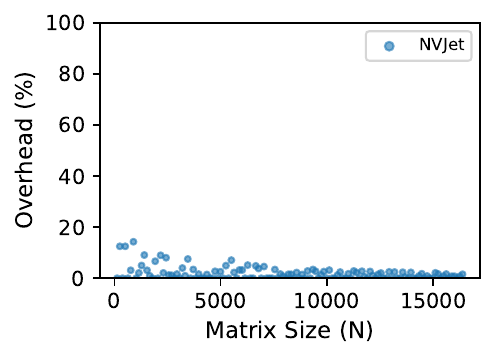}
    \caption{H100 FP8}
  \end{subfigure}
  \\[6pt]
  \begin{subfigure}[t]{0.24\textwidth}
    \includegraphics[width=\linewidth]{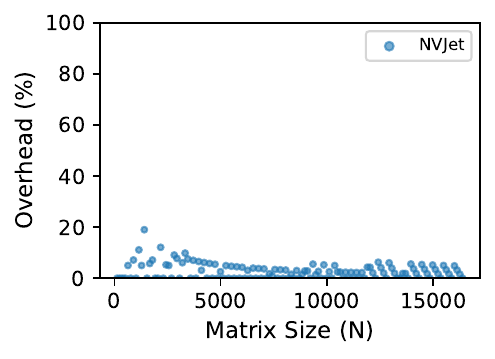}
    \caption{GB200 FP16}
  \end{subfigure}\hfill
  \begin{subfigure}[t]{0.24\textwidth}
    \includegraphics[width=\linewidth]{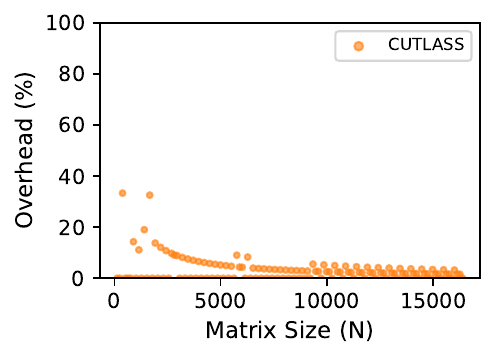}
    \caption{GB200 TF32}
  \end{subfigure}\hfill
  \begin{subfigure}[t]{0.24\textwidth}
    \includegraphics[width=\linewidth]{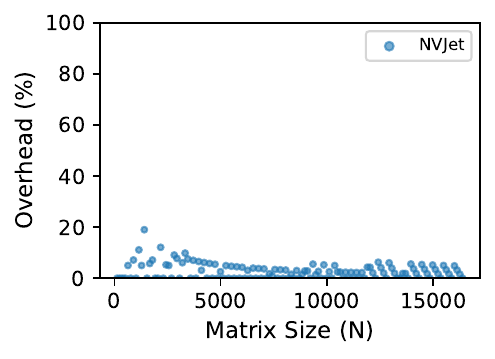}
    \caption{GB200 FP8}
  \end{subfigure}\hfill
  \begin{subfigure}[t]{0.24\textwidth}
    \includegraphics[width=\linewidth]{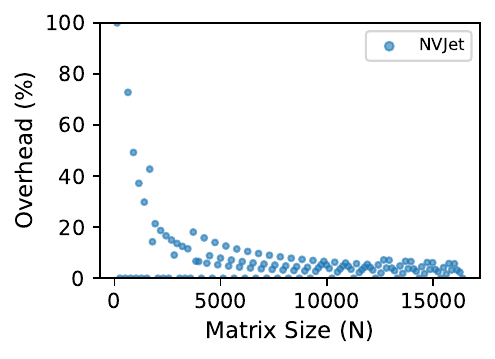}
    \caption{GB200 NVFP4}
  \end{subfigure}
  \caption{FLOP overhead for square matrices due to tiling and cuBLAS
  kernel selection.}
  \label{fig:flop-overhead}
\end{figure*}

\subsection{Deriving the MFU Estimator}

We use Tensor Pipe Activity (TPA) as the base signal,
which reports the fraction of cycles during which the GPU is executing
Tensor Core floating-point instructions across all GPU generations:
\[
  \mathrm{TPA}
  \;=\;
  \frac{\text{Cycles GPU is executing Tensor instructions}}{\text{Total Cycles}}
\]
TPA is a hardware-averaged counter: the GPU accumulates active and total
cycle counts over the collection window and reports their ratio, so a single
readout already reflects the true mean activity over that interval.
Thus, the per-cycle approximation is given by:
\[
  \mathrm{TPA}
  \;\approx\;
  \frac{\text{Actual FLOPs/cycle}}{\text{Peak FLOPs/cycle}}.
\]

TPA is measured in \emph{cycles}, but MFU is a throughput ratio in
\emph{FLOPs per second}. The link between the two is the SM clock
frequency. The theoretical peak FLOP/s of a GPU is defined at the
maximum boost clock $f_{\mathrm{SM}}^{\max}$. When the GPU runs at a lower clock
$f_{\mathrm{SM}} < f_{\mathrm{SM}}^{\max}$, every cycle still
delivers the same FLOPs per cycle, but fewer cycles elapse per second,
so the realized FLOP/s is reduced proportionally. Multiplying the
per-cycle ratio by the clock ratio therefore converts from a cycle-domain
metric to a time-domain one:
\[
  \frac{\text{Actual FLOPs/cycle}}{\text{Peak FLOPs/cycle}}
  \;\times\;
  \frac{f_{\mathrm{SM}}}{f_{\mathrm{SM}}^{\max}}
  \;=\;
  \frac{\text{Actual FLOP/s}}{\text{Peak FLOP/s}}
  \;\approx\;
  \mathrm{MFU}.
\]

In practice, $f_{\mathrm{SM}}$ is not constant. The GPU's power and
thermal management continuously adjusts the SM clock in response to
workload intensity, power draw, and GPU temperature---even during a
sustained single-kernel workload, the clock can fluctuate by hundreds of
megahertz multiple times per second. Unlike TPA,
which is hardware-averaged over the collection window, the SM clock
reported by the hardware (\texttt{DCGM\_FI\_DEV\_SM\_CLOCK}) is an
\emph{instantaneous} point sample. A single readout can therefore
over- or underestimate the true mean clock for that interval,
introducing noise into the OFU estimate. Averaging OFU over many collection points
mitigates this sampling bias, but characterizing the residual
error is critical to establishing bounded accuracy for any
hardware-counter-based metric.

\subsection{Overall FLOP Utilization (OFU)}
We refer to this hardware-derived estimate as \emph{Overall FLOP Utilization}
(OFU):
\begin{equation}
  \mathrm{OFU}
  \;=\;
  \mathrm{TPA}
  \;\times\;
  \frac{f_{\mathrm{SM}}}{f_{\mathrm{SM}}^{\max}}
  \label{eq:ofu}
\end{equation}
We use the term \emph{Overall} FLOP Utilization to distinguish it from
application-level MFU: OFU captures \emph{all} FLOPs executed at the Tensor Core hardware level, not only
the forward and backward pass FLOPs that application-level MFU typically
counts; and OFU is
\emph{precision-agnostic}, since the hardware counter measures Tensor Core
activity regardless of numeric format.

\section{Characterizing OFU Properties}
\label{sec:characterization}

Before deploying OFU in production, we must understand its behavior
across hardware configurations, numeric formats, and operating
conditions. We examine five properties:
\emph{tile quantization and cuBLAS kernel selection}, where fixed-size
tiles and zero-padding cause the hardware to execute more FLOPs than the
theoretical $2MNK$ cost;
\emph{floating-point precision}, where we validate that OFU correctly
tracks utilization across numeric formats (FP16, TF32, FP8, NVFP4) and
GPU architectures (H100, GB200);
\emph{SM clock sampling noise}, where the instantaneous clock sample
can diverge from the true mean, widening the confidence interval of any
single OFU reading;
\emph{theoretical peak FLOPs}, where Tensor Core pipelines may operate
at a different maximum clock frequency than the SM boost clock; and
\emph{non-tensor undercounting}, where OFU monitors only the Tensor
Core pipeline.

\subsection{Tile Quantization and cuBLAS Kernel Selection}
\label{sec:gemm_validation}

Tile quantization is a software constraint imposed by GEMM kernels to
maximize Tensor Core throughput. Because virtually all AI training computation
reduces to matrix multiplications (attention, linear layers,
convolutions)~\cite{ivanov2021datamovementallneed},
GEMM is the natural workload for isolating this error. For $C = A \times B$ where
$A$ is $M\!\times\!K$ and $B$ is $K\!\times\!N$, the theoretical cost is
exactly $2MNK$ FLOPs. Because the workload is fully specified, any
discrepancy between OFU and the true utilization must originate in the
hardware execution---not in uncertainty about what the application is
computing.

We profiled matrix multiplications on H100 and GB200 GPUs using
NVIDIA Nsight Compute (NCU) inside the
\mbox{\texttt{nvcr.io/nvidia/pytorch:25.11-py3}} container. For FP16 and TF32, we used PyTorch \texttt{torch.matmul}; for FP8, we used \texttt{torch.\_scaled\_mm}; for NVFP4
on GB200, we used an internal matrix multiplication benchmark, as
PyTorch does not yet have full native NVFP4 GEMM support at the time of writing.
For each $(M,K,N)$ triple we executed a
single matrix multiplication, collected the precision-specific
tensor-op counter, and computed the FLOP overhead, the fraction of
extra FLOPs the hardware executes beyond the theoretical $2MNK$ due to
tile padding:
\begin{equation}
  \text{Overhead} =
  \frac{\text{FLOPs}_{\text{profiled}} - 2MNK}{2MNK}\times 100\%
\end{equation}
We tested FP16, TF32, FP8, and NVFP4 using square matrices from $N\!=\!128$ to
$N\!=\!16384$ in increments of 128 (tensor-core-friendly
alignment~\cite{nvidia_matmul_perf}), plus ${\sim}$1000 square
matrices with randomly chosen dimensions (not necessarily multiples of 128).

Fig.~\ref{fig:flop-overhead} shows the measured overhead for H100 and GB200.
Two patterns are clear: overhead decreases with matrix size as padding
waste becomes a smaller fraction of total work, and overhead varies by
precision and GPU due to differing tile sizes selected by
cuBLAS~\cite{nvidia_cublas_docs}.

For well-aligned matrices (multiples of 128) with $N \geq 4096$, the
maximum overhead observed was ${\sim}$9\% across both GPUs and all
precisions, with means of 2--3\%. For non-aligned matrices (Fig.~\ref{fig:flop-overhead}b; GB200 random omitted as
it exhibits a similar pattern), overhead at $N \geq 4096$ reached up to
${\sim}$12\%, though the mean remained around 5\%. At small sizes ($N < 512$, rarely used in large-model training), overhead
can exceed 50\% due to severe tile quantization.

As shown in Fig.~\ref{fig:flop-overhead}, FP16, FP8, and NVFP4 exhibit nearly
identical overhead curves on GB200: all are routed
exclusively to nvJet kernels and converge to approximately 2--4\% overhead
for matrices above $N = 4096$. TF32 is a notable outlier---cuBLAS
selects XMMA and CUTLASS kernels instead of nvJet, producing
systematically higher overhead (up to 33\% at small sizes) that converges
more slowly. This suggests that cuBLAS kernel selection, a software optimization, is another significant factor in the tile-quantization overhead.

cuBLAS employs a comprehensive set of heuristics that optimize kernel dispatch based on matrix shape, kernel implementation, precision GPU architecture, clocks, and available pipelines ~\cite{nvmmh}. These heuristics can spread computation across multiple execution pipelines in ways that
do not correspond to the user's chosen datatype---for example, TF32
operations may be dispatched to kernels that heavily utilize the FP16
(HMMA) pipeline. This observation further motivates a hardware-level
metric like OFU: optimization opportunities arise not only from
application-level choices and hardware capabilities, but also from
intermediate library layers whose behavior is opaque to both the user
and the training framework. Application-level MFU, which derives FLOPs
from model architecture, cannot capture these library-level effects. OFU,
by measuring what the GPU actually executes, reflects the true
utilization regardless of how cuBLAS maps the workload to hardware
pipelines.

\begin{figure}[!htb]
  \centering
  \begin{subfigure}[t]{0.49\linewidth}
    \centering
    \includegraphics[width=\linewidth]{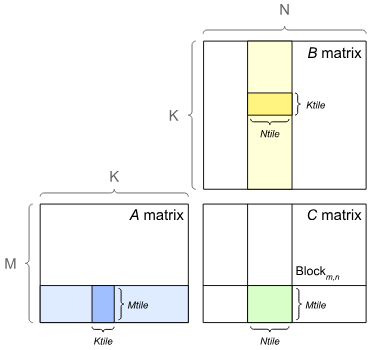}
    \caption{Tiled matrix multiplication.}
    \label{fig:tile-matmul}
  \end{subfigure}\hfill
  \begin{subfigure}[t]{0.49\linewidth}
    \centering
    \includegraphics[width=\linewidth]{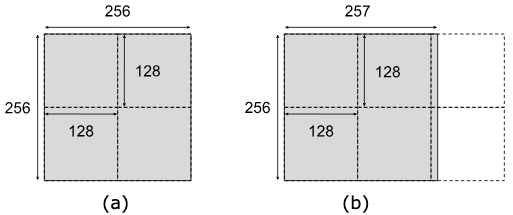}
    \caption{Tile quantization impact.}
    \label{fig:tile-quant}
  \end{subfigure}
  \caption{Tile-quantization overhead in GEMM
  execution~\cite{nvidia_matmul_perf}.}
  \label{fig:tiling}
\end{figure}

GPU GEMM kernels partition the output matrix $C$ into rectangular tiles
assigned to thread blocks~\cite{nvidia_matmul_perf}
(Fig.~\ref{fig:tile-matmul}). When $M$, $N$, or $K$ does not divide evenly
into the tile dimensions $(T_M, T_N, T_K)$, the last tiles are zero-padded in shared memory
and computed in full (Fig.~\ref{fig:tile-quant}), yielding effective
dimensions:
\begin{equation}
  M_{\text{eff}} = \left\lceil \frac{M}{T_M} \right\rceil T_M,\quad
  N_{\text{eff}} = \left\lceil \frac{N}{T_N} \right\rceil T_N,\quad
  K_{\text{eff}} = \left\lceil \frac{K}{T_K} \right\rceil T_K
\end{equation}
The actual FLOPs executed are $2\,M_{\text{eff}}\,N_{\text{eff}}\,
K_{\text{eff}} \geq 2MNK$. This is known as \emph{tile
quantization}~\cite{nvidia_matmul_perf}.

The cuBLAS heuristics (which among other heuristics leverages nvMatmulHeuristics ~\cite{nvidia_cutlass_heuristics}) select from several
kernel families depending on matrix shape and precision. The primary families
observed in our experiments are:

\begin{itemize}
  \item \textbf{nvJet}: NVIDIA's proprietary high-performance GEMM kernels, which seem to efficiently expose a very large number of tile count, precisions and fused epilogues. Selected for most well-aligned shapes that can leverage TMA.
  \item \textbf{XMMA}: CUDA C++ template-based kernels focused on Cooperative Thread Array (CTA)-level decomposition, used by cuBLAS and cuDNN.
  \item \textbf{CUTLASS 2/3}: Open-source GEMM templates; CUTLASS~2 lacks
    Cooperative Grid Array (CGA) support and is typically selected for small or poorly aligned matrices.
\end{itemize}

Modern kernels (nvJet, XMMA, CUTLASS~3) use CGAs~\cite{nvidia_hopper_wp},
which group $(C_M, C_N)$ thread blocks into clusters that share distributed
shared memory across SMs. This introduces a \emph{two-level} tiling
hierarchy: at the first level, each thread block computes one
$T_M \times T_N$ output tile, rounding the matrix dimensions up to tile
boundaries as described above; at the second level, thread blocks are
grouped into $C_M \times C_N$ clusters, and the number of tiles must itself
be rounded up to a whole number of clusters. The effective dimension
therefore undergoes two successive ceiling operations:
\begin{equation}
  M_{\text{eff}} =
    \left\lceil\frac{\lceil M/T_M\rceil}{C_M}\right\rceil
    \cdot C_M \cdot T_M
\end{equation}
(and analogously for $N$). When $C_M > 1$, a matrix that fits exactly into
an integer number of tiles can still incur padding at the cluster level,
adding an extra $C_M - 1$ tiles of waste in the worst case. For nvJet kernels, the tile dimensions and CGA configuration are encoded in the kernel name (e.g.\
\texttt{nvjet\_sm90\_hsh\_256x160\_64x4\_2x1}), enabling a closed-form FLOP
prediction that matched NCU measurements to within $<$\,1000~FLOPs for all
tested cases. For XMMA and CUTLASS kernels the $K$-dimension tiling and CGA
configuration are not statically visible, so exact prediction requires
runtime introspection.

\begin{figure}[!t]
  \centering
  \includegraphics[width=\linewidth]{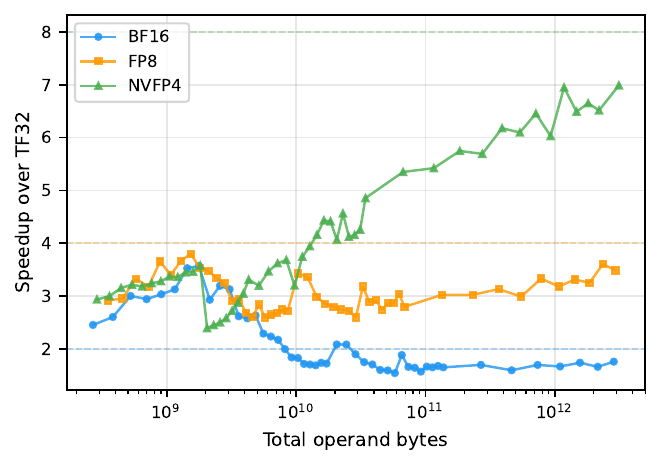}
  \caption{Throughput speedup over TF32 on GB200.}
  \label{fig:precision_speedup}
\end{figure}

\subsection{Floating-Point Precision}
\label{sec:cross_precision}

Precision format affects realized throughput relative to the theoretical
peak. To characterize this, we profiled square GEMMs on a GB200 GPU
across increasing matrix sizes in BF16, FP8, and NVFP4, with $N$
ranging up to $18{,}432$ for TF32/BF16, $24{,}576$ for FP8, and
$32{,}768$ for NVFP4, measuring the ratio of achieved throughput (FLOP/s) relative to TF32
(Fig.~\ref{fig:precision_speedup}). At large matrix sizes the curves
approach their theoretical speedups: $2\times$ for BF16, $4\times$ for
FP8, and $8\times$ for NVFP4, with BF16 converging most cleanly because
$2\times$ is the smallest multiplier. Lower precisions show
progressively more deviation, partly due to scaling-factor (SF) overhead
in block-scaled formats: FP8 requires one 512-byte SF block per
$128\times128$ input tile, while NVFP4 requires one per
$128\times64$ input tile. Considering typical FP8 tiles are $128\times256\times128$ and NVFP4 are $128\times256\times256$ we quadruple the SF overhead, going from 3 to 12 SF blocks per tile. 
NVFP4's speedup falls below $8\times$ at small matrix sizes where SF
bookkeeping dominates, recovering as matrix dimensions grow.
OFU-derived speedup, computed as $(\text{OFU}_p \times \text{Peak}_p) /
(\text{OFU}_{\text{TF32}} \times \text{Peak}_{\text{TF32}})$, closely
tracks the measured curves: at $N \geq 4096$, OFU-derived speedups of
$1.85\times$, $3.51\times$, and $6.75\times$ for BF16, FP8, and NVFP4
agree with the measured $1.78\times$, $3.27\times$, and $6.10\times$,
confirming that OFU correctly captures precision-dependent throughput
scaling.

\subsection{SM Clock Sampling Noise}
\label{sec:clock_sampling}

Clock sampling introduces noise into OFU estimates. OFU is computed
from two hardware counters, tensor core
activity and SM clock frequency, that are polled at discrete intervals.
Because the SM clock reported by the hardware is an instantaneous sample rather
than a hardware-averaged value (unlike tensor pipe activity, which is
averaged
over the collection window), coarser scrape intervals introduce sampling
noise into the OFU estimate. For example, during a sustained
$16384 \times 16384$ BF16 GEMM on an H100, the SM clock sampled at
1\,kHz via Nsight Systems fluctuates between ${\sim}$1{,}201\,MHz and
${\sim}$1{,}558\,MHz (mean 1{,}352\,MHz, std 32\,MHz), driven by
power and thermal management.

To quantify the impact on OFU, we collected both counters at 1-second
intervals, the minimum supported by \texttt{nvidia-smi dmon}, over
3{,}000~seconds of sustained FP16 matrix multiplication on a GB200 GPU,
then subsampled at coarser intervals (5--30\,s) and measured the
deviation from the 1-second baseline. (DCGM supports collection
intervals as low as 100\,ms, which would further reduce sampling error.) Three steady-state matrix sizes ($N = 4096$, $8192$,
$16384$) and an alternating workload ($16384 \leftrightarrow 4096$,
switching every 10\,s) were tested.
Table~\ref{tab:clock-sampling-error} reports the results.

\begin{table}[t]
\centering
\caption{Errors from clock frequency sampling rates.}
\label{tab:clock-sampling-error}
\resizebox{\columnwidth}{!}{%
\begin{tabular}{c|cc|cc|cc|cc}
\toprule
\textbf{Int.} & \multicolumn{2}{c|}{$N{=}4096$} & \multicolumn{2}{c|}{$N{=}8192$} & \multicolumn{2}{c|}{$N{=}16384$} & \multicolumn{2}{c}{Alt.} \\
(s) & $\sigma$ & 95\% & $\sigma$ & 95\% & $\sigma$ & 95\% & $\sigma$ & 95\% \\
\midrule
5  & 0.01 & $\pm$0.01 & 0.03 & $\pm$0.05 & 0.03 & $\pm$0.07 & 0.01 & $\pm$0.03 \\
10 & 0.02 & $\pm$0.04 & 0.07 & $\pm$0.13 & 0.09 & $\pm$0.18 & 0.03 & $\pm$0.06 \\
20 & 0.04 & $\pm$0.07 & 0.08 & $\pm$0.15 & 0.03 & $\pm$0.06 & 0.08 & $\pm$0.15 \\
30 & 0.02 & $\pm$0.03 & 0.09 & $\pm$0.18 & 0.11 & $\pm$0.22 & 0.09 & $\pm$0.17 \\
\bottomrule
\end{tabular}}
\end{table}

Even at 30-second intervals (${\sim}$100~samples over a 50-minute
window), the 95\% confidence interval remains below $\pm$0.22
percentage points---negligible relative to the OFU values themselves
(${\sim}$55\%). At 5-second intervals the bound drops below
$\pm$0.07 percentage points. Sampling noise is therefore not a material
source of error for production OFU measurements, provided the collection
window spans at least several minutes.

A practical constraint is that the DCGM hardware counter for tensor pipe
activity averages over at most 30-second windows. Collecting at
intervals longer than 30\,seconds yields an average of averages rather
than a true window average, compounding estimation error. Thus, a collection interval should be at most 30\,seconds.

\subsection{Theoretical Peak FLOPs and Tensor Core Clock Domains}
\label{sec:theoretical_flops}

Computing OFU requires a correct peak TFLOP/s denominator, which depends
on the clock frequency of the Tensor Core pipeline.
The theoretical peak throughput of a GPU is determined by three
architectural parameters: the number of Streaming Multiprocessors (SMs),
the number of floating-point operations each SM can perform per clock cycle
on the relevant execution pipeline, and the maximum clock frequency of
that pipeline:
\begin{equation}
  \text{Peak TFLOP/s} = \frac{\text{SMs} \times \text{FLOPs/cycle/SM}
    \times f^{\max}}{10^{12}}
  \label{eq:peak_tflops}
\end{equation}

A subtlety arises because Tensor Core pipelines do not necessarily run
at the same maximum clock frequency as the rest of the
SM~\cite{nvidia_h100_wp}. On the H100~SXM, Tensor Core operations in
lower-precision formats (FP8, FP16, BF16, TF32) boost to a maximum of
1{,}830~MHz, whereas the SM boost clock is 1{,}980~MHz. FP32 and FP64
operations, including FP32 and FP64 Tensor Core/Non Tensor Core instructions, run at
the full 1{,}980~MHz SM clock.

This distinction is critical for computing correct peak throughput
values. Using the primary clock for Tensor Core precisions on H100:
\begin{equation}
\begin{split}
  \text{Peak}_{\text{H100, FP16}}
    &= \frac{132 \times 4{,}096 \times 1{,}830 \times 10^{6}}{10^{12}} \\
    &= 989.4 \;\text{TFLOP/s}
\end{split}
\end{equation}
which agrees with the published specification of
989~TFLOP/s~\cite{nvidia_h100_product_page}. The remaining Tensor Core
precisions scale proportionally from this base rate:
\begin{itemize}
  \item \textbf{FP8}: $2 \times 989 = 1{,}978$~TFLOP/s.
  \item \textbf{TF32}: $989\,/\,2 = 494.5$~TFLOP/s.
\end{itemize}

For the GB200, no public documentation currently specifies a separate
Tensor Core clock frequency. Using the published SM boost clock of
2{,}062~MHz as the Tensor Core frequency:
\begin{equation}
\begin{split}
  \text{Peak}_{\text{GB200, FP16}}
    &= \frac{148 \times 8{,}192 \times 2{,}062 \times 10^{6}}{10^{12}} \\
    &= 2{,}499.9 \;\text{TFLOP/s}
\end{split}
\end{equation}
which matches the published specification of
2{,}500~TFLOP/s~\cite{nvidia_gb200_product_page}.

\subsection{Non-Tensor Undercounting}
\label{sec:undercounting}

OFU monitors only the Tensor Core pipeline, excluding CUDA-core work
(activations, normalization, softmax). This omission is negligible:
matrix multiplications account for 99.8\% of total FLOPs in a
transformer encoder
layer~\cite{ivanov2021datamovementallneed}. Standard MFU definitions---PaLM~\cite{chowdhery2022palm},
Megatron-LM~\cite{korthikanti2022reducing}, and the OpenAI scaling
laws~\cite{kaplan2020scaling}---follow the same convention, deriving
FLOPs exclusively from matrix-multiplication terms.

\section{Practical Accuracy}
\label{sec:evaluation}

In this section, we evaluate OFU's practical accuracy. We first apply
tile-quantization corrections to controlled GEMM workloads, establishing
bounded accuracy on fully specified workloads. We then compare OFU
against application-reported MFU on 608 production training jobs,
measuring correlation and surfacing cases where the two metrics diverge.

\subsection{Predicting MFU from Hardware Counters}
\label{sec:mfu_prediction}

Using the tile-quantization corrections from
Section~\ref{sec:gemm_validation} and the Tensor Core clock frequencies
from Section~\ref{sec:theoretical_flops}, we evaluate how accurately OFU
tracks application-level MFU on sustained matrix multiplications.

We compare three quantities for each matmul:
\begin{itemize}
  \item \textbf{OFU} (unadjusted), as defined in \eqref{eq:ofu}, computed
    from Nsight Systems GPU metrics sampled at 10\,kHz.
  \item \textbf{Adjusted OFU}, which corrects OFU for tile-quantization
    overhead (Section~\ref{sec:gemm_validation}):
    \begin{equation}
      \text{OFU}_{\text{adj}} = \text{OFU} \times
      \frac{\text{FLOPs}_{\text{theoretical}}}{\text{FLOPs}_{\text{profiled}}}
      = \text{OFU} \times
      \frac{2MNK}{\text{FLOPs}_{\text{NCU}}}
    \end{equation}
  \item \textbf{App MFU} (ground truth), computed as measured TFLOP/s divided
    by the architecturally derived peak TFLOP/s for the given precision
    (Section~\ref{sec:theoretical_flops}).
\end{itemize}

For each GPU and precision we profiled 500 random $(M,K,N)$ matrix
multiplications, where each dimension
was a random multiple of~16. Each matmul ran for 5~minutes under Nsight
Systems profiling, collecting both application throughput and hardware
counters (tensor pipe activity and SM clock frequency) at 10\,kHz. A
single-iteration NCU pass was then run to measure the actual FLOP count for
the tile-quantization correction. For FP16 and TF32 we used
PyTorch \texttt{torch.matmul} and for FP8 we used \texttt{torch.\_scaled\_mm}, inside the
\texttt{nvcr.io/nvidia/pytorch:25.11-py3} container; for NVFP4 on GB200 NVL
we used an internal matrix multiplication benchmark;
NVFP4 dimensions were restricted to multiples of~128. We profiled 500 random
$(M,K,N)$ GEMMs per configuration: FP16, TF32, and FP8 on H100 SXM, and
FP16, TF32, FP8, and NVFP4 on GB200 NVL, using the architecturally
derived peak TFLOP/s from Section~\ref{sec:theoretical_flops}.

\begin{figure*}[t]
  \centering
  \begin{subfigure}[t]{0.48\textwidth}
    \centering
    \includegraphics[width=\linewidth]{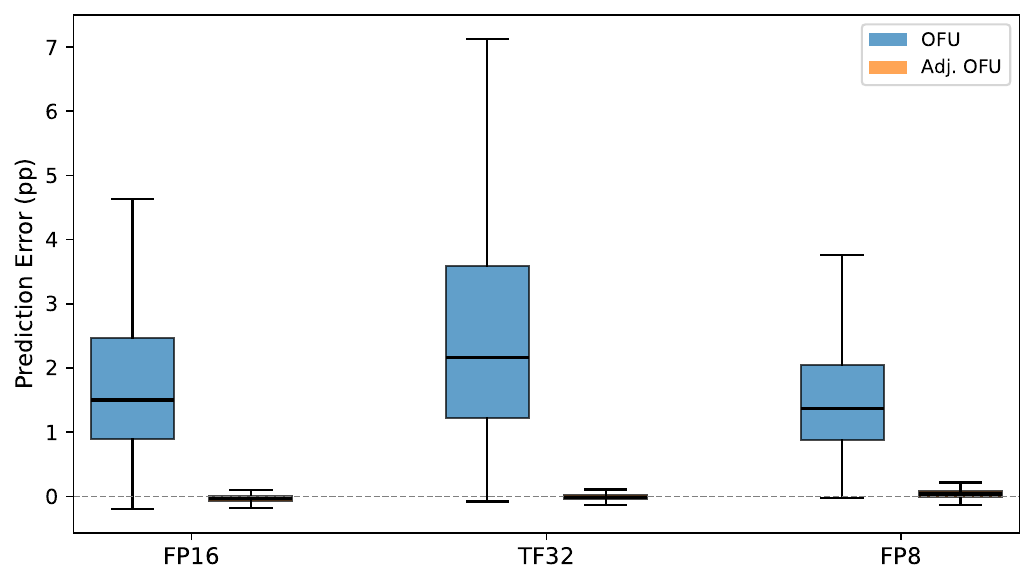}
    \caption{H100}
    \label{fig:h100-pred-error}
  \end{subfigure}\hfill
  \begin{subfigure}[t]{0.48\textwidth}
    \centering
    \includegraphics[width=\linewidth]{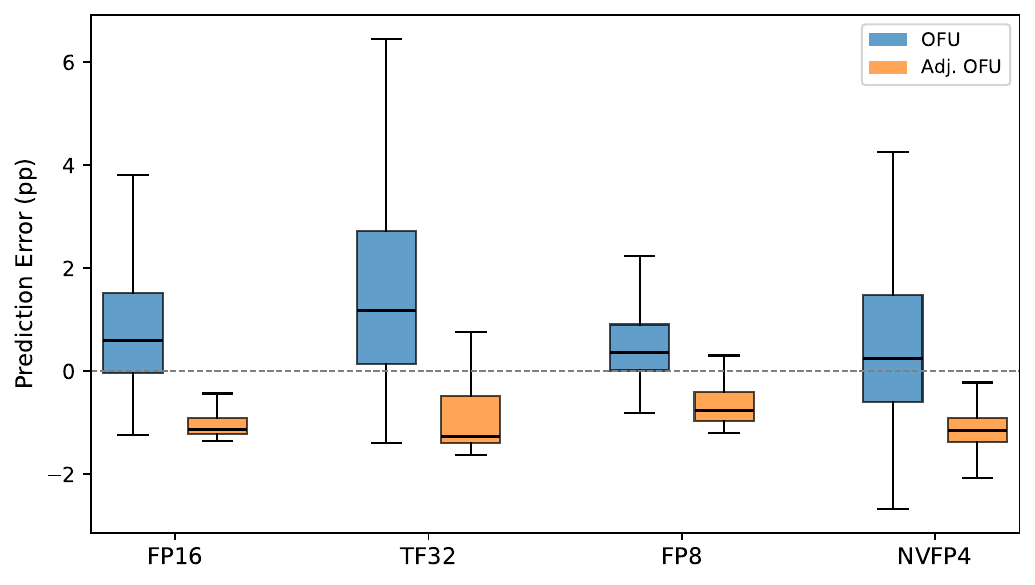}
    \caption{GB200}
    \label{fig:gb200-pred-error}
  \end{subfigure}
  \caption{OFU prediction error.}
  \label{fig:pred-error}
\end{figure*}

Fig.~\ref{fig:h100-pred-error} and Fig.~\ref{fig:gb200-pred-error} show the
distribution of prediction error (estimate -- App~MFU, in percentage
points) across all 500 matmuls per configuration. Table~\ref{tab:mfu-prediction-stats}
provides summary statistics, where the mean absolute error (MAE) is
defined as:
\begin{equation}
  \text{MAE} = \frac{1}{n} \sum_{i=1}^{n}
    \left| \text{OFU}_i - \text{App\,MFU}_i \right|
\end{equation}

\begin{table}[t]
\centering
\caption{Prediction accuracy.}
\label{tab:mfu-prediction-stats}
\small
\begin{tabular}{llrrrr}
\toprule
GPU & Prec & Estimator & MAE & $\leq$2\,pp & $\leq$5\,pp \\
\midrule
\multirow{2}{*}{H100}
  & \multirow{2}{*}{FP16} & OFU     & 1.90 & 64\% & 96\% \\
  &                        & Adj OFU & 0.06 & 100\% & 100\% \\
\midrule
\multirow{2}{*}{H100}
  & \multirow{2}{*}{TF32} & OFU     & 3.46 & 44\% & 86\% \\
  &                        & Adj OFU & 0.50 & 99\% & 99\% \\
\midrule
\multirow{2}{*}{H100}
  & \multirow{2}{*}{FP8}  & OFU     & 1.58 & 73\% & 99\% \\
  &                        & Adj OFU & 0.07 & 100\% & 100\% \\
\midrule
\multirow{2}{*}{GB200}
  & \multirow{2}{*}{FP16} & OFU     & 1.08 & 86\% & 99\% \\
  &                        & Adj OFU & 1.04 & 100\% & 100\% \\
\midrule
\multirow{2}{*}{GB200}
  & \multirow{2}{*}{TF32} & OFU     & 2.10 & 65\% & 88\% \\
  &                        & Adj OFU & 1.03 & 100\% & 100\% \\
\midrule
\multirow{2}{*}{GB200}
  & \multirow{2}{*}{FP8}  & OFU     & 0.64 & 96\% & 100\% \\
  &                        & Adj OFU & 0.70 & 100\% & 100\% \\
\midrule
\multirow{2}{*}{GB200}
  & \multirow{2}{*}{NVFP4} & OFU     & 1.21 & 87\% & 98\% \\
  &                         & Adj OFU & 1.15 & 95\% & 100\% \\
\bottomrule
\end{tabular}
\end{table}

Raw OFU consistently overestimates App~MFU by
1--2~percentage points across both GPUs and all precisions, as expected from the
tile-quantization overhead identified in Section~\ref{sec:gemm_validation}.
After NCU correction, Adjusted~OFU centres near zero with substantially
reduced variance.

On H100, Adjusted~OFU achieves $\leq$2~percentage-point error for 99--100\% of matmuls
across all three precisions, with mean absolute error under 0.6~percentage points. On
GB200, Adjusted~OFU achieves $\leq$2~percentage-point error for 95--100\% of matmuls and
$\leq$5~percentage points for 100\% across all four precisions (FP16, TF32, FP8, NVFP4),
though a small systematic underestimate of
${\sim}$1~percentage point remains, likely from sampling overhead in the 10\,kHz
hardware counters.

For pure GEMM workloads, OFU predicts application-level MFU to within
1--3~percentage points without any model-specific information. The NCU-adjusted estimator
further tightens this to $\leq$2~percentage points for all tested configurations.

\subsection{Production Workloads Validation}
\label{sec:production_validation}

We evaluated OFU as a proxy for MFU across 608 production
training jobs on H100 GPUs at a commercial GPU cluster
(August 27--September 10, 2025), run by an internal research group using Megatron-LM. Jobs ranged from 8 to 5,888 GPUs across 80 distinct
configurations from 26 users.
 
MFU was sourced from \toolLogger, computed from
Megatron-LM's reported total FLOPs and training loop wall-clock time:
\begin{equation}
  \text{MFU} = \frac{\texttt{train\_tflop} \times \texttt{gpu\_count}}{989} \times 100\%
  \label{eq:mfu}
\end{equation}
where 989~TFLOP/s is the H100 BF16 Tensor Core
peak~\cite{nvidia_h100_product_page} (see
Section~\ref{sec:theoretical_flops}).

DCGM telemetry was scraped via Prometheus at 30-second intervals, aligned to
each job's training window. OFU was computed as:
\begin{equation}
  \text{OFU} = \mathrm{mean}\!\left(
    \texttt{\small PIPE\_TENSOR\_ACTIVE} \times
    \frac{\texttt{\small SM\_CLOCK}}{1830} \times 100
  \right)
\end{equation}
averaged across all GPUs and time samples, where 1830~MHz is the H100
Tensor Core maximum clock frequency
(Section~\ref{sec:theoretical_flops}).
 
\begin{figure}[t]
  \centering
  \includegraphics[width=0.75\columnwidth]{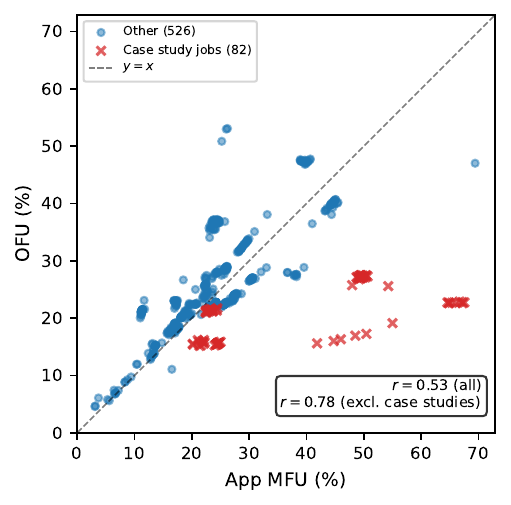}
  \caption{App MFU vs.\ OFU for 608 production training jobs.}
  \label{fig:mfu-vs-ofu}
\end{figure}

Across all 608 jobs, MFU and OFU show a moderate positive
correlation (Pearson $r = 0.53$). Mean MFU was $25.1\% \pm 10.9\%$ versus
mean OFU of $25.0\% \pm 8.3\%$. Mean absolute error was
$6.2\%$. Of all jobs, 79.4\% fell within 10\% absolute error, while 6.7\%
exceeded 20\% error. Fig.~\ref{fig:mfu-vs-ofu} shows the per-job
relationship; most jobs cluster near the $y = x$ line, with the 288-GPU MoE
group as a clear outlier.

Table~\ref{tab:scale-analysis} summarizes results by GPU count. Agreement
improves substantially at large scale: jobs with $\geq$768 GPUs consistently
achieve sub-5\% absolute error. The most significant outlier is the 288-GPU
group, which exhibits a mean absolute error of 18.0\%, driven by a specific
architectural issue described below.

\subsection{Detecting Production FLOPs Miscalculations}
\label{sec:case_studies}

To illustrate the kinds of production problems our techniques help
detect and remediate, we examined jobs where MFU and OFU diverged
most significantly.
This analysis surfaced two distinct FLOPs miscalculations in Megatron-LM. Excluding the affected
82 jobs improves overall correlation from $r = 0.53$ to $r = 0.78$, and
reduces the fraction of jobs exceeding 10\% absolute error from 21.8\% to
16.7\%.
 
A representative 288-GPU job training a 16B-parameter DeepSeek-style MoE where activations were down-projected from hidden
dimension 2048 to latent dimension 512 before expert routing.
Megatron-LM's FLOPs counter incorrectly assumed experts operated at the full
hidden dimension (2048), inflating the reported FLOPs by a factor of
${\sim}3\times$. This produced an application-reported MFU of 54.27\% against
OFU of 25.58\% (relative error: 112.2\%).
Correcting the FLOPs count to account for the down- and up-projections reduced
the reported MFU to 18.45\%, cutting the relative error to 27.9\%.
 
A second miscalculation affected hybrid MoE jobs training an 8B-parameter model on 300B tokens
that interleave attention, Mamba~\cite{gu2024mamba}, dense MLP, and sparse MoE layers.  The Megatron-LM branch used for
these experiments did not support hybrid architectures in its FLOPs
counter. As a result, every layer's FLOPs were counted as if it were a self-attention and dense MLP layer
cost.  This inflated the reported FLOPs, producing
MFU of 24.51\% against OFU of 15.56\% (relative error: 57.5\%).
After the FLOPs function was updated with per-layer-type accounting,
subsequent runs of a similar architecture (1,536 GPUs) reported MFU of
17.8--18.0\% versus OFU of 18.5--18.7\% (relative error 3--4\%).

Together, these case studies indicate that significant divergence between
OFU and application-reported MFU consistently traced back to incorrect
FLOPs calculations in the training framework rather than OFU measurement
error. This is expected: application-level MFU depends on manually
derived FLOPs counts that are brittle for novel architectures, whereas
OFU is computed directly from hardware counters with no model-specific
assumptions.

\begin{table}[t]
\centering
\caption{Absolute error between MFU and OFU.}
\label{tab:scale-analysis}
\small
\begin{tabular}{rrrr}
\toprule
GPUs & Jobs & MFU (\%) & Abs Err (\%) \\
\midrule
8    & 6   & 28.7 $\pm$ 6.9  & 7.5 $\pm$ 3.9 \\
16   & 48  & 23.8 $\pm$ 3.3  & 12.2 $\pm$ 2.0 \\
64   & 52  & 23.6 $\pm$ 2.5  & 2.2 $\pm$ 2.4 \\
128  & 48  & 24.3 $\pm$ 8.7  & 4.5 $\pm$ 2.9 \\
256  & 76  & 20.1 $\pm$ 12.6 & 9.1 $\pm$ 4.9 \\
288  & 65  & 40.1 $\pm$ 16.3 & 18.0 $\pm$ 14.4 \\
512  & 144 & 23.9 $\pm$ 5.6  & 3.6 $\pm$ 2.2 \\
736  & 11  & 24.2 $\pm$ 0.4  & 3.6 $\pm$ 0.1 \\
768  & 57  & 16.9 $\pm$ 4.1  & 1.2 $\pm$ 0.7 \\
1024 & 49  & 35.0 $\pm$ 9.1  & 4.1 $\pm$ 0.7 \\
1536 & 10  & 12.4 $\pm$ 2.3  & 0.3 $\pm$ 0.2 \\
2944 & 33  & 24.0 $\pm$ 3.7  & 2.6 $\pm$ 0.3 \\
5888 & 9   & 13.6 $\pm$ 0.1  & 1.5 $\pm$ 0.2 \\
\bottomrule
\end{tabular}
\end{table}

\section{Operational Experiences}
\label{sec:operational}

OFU has been integrated at multiple levels of a large-scale GPU fleet
infrastructure---from per-job dashboards visible to individual
researchers, to cluster-wide resilience and goodput services that flag inefficiencies
and drive optimization. In each deployment, OFU discovered problems that were hidden
due to the lack of instant visibility into floating-point behavior with
well-defined accuracy properties. This section
describes how OFU is operationalized in practice and presents the
problems it uncovered.

\subsection{Embodied Agent Training}
\label{sec:embodied}

To evaluate whether OFU is useful in practice, we operationalized the metric
for an internal research lab that develops foundation models for robotic
embodied agents. We integrated OFU into
\toolOrch~\cite{nvidia_osmo}, the lab's Kubernetes-native orchestration platform for
Physical AI workloads, so that researchers could monitor GPU utilization at
the job level without deriving model-specific FLOPs counts.

For each training job managed by \toolOrch, we compute OFU from DCGM metrics
and display it as a time-series dashboard, both per GPU individually and
as a job-level aggregate. Separately, the lab's training
infrastructure optionally runs a few iterations with PyTorch Profiler at job
start and uploads the trace to S3 for later analysis.

\begin{figure}[t]
  \centering
  \includegraphics[width=\columnwidth]{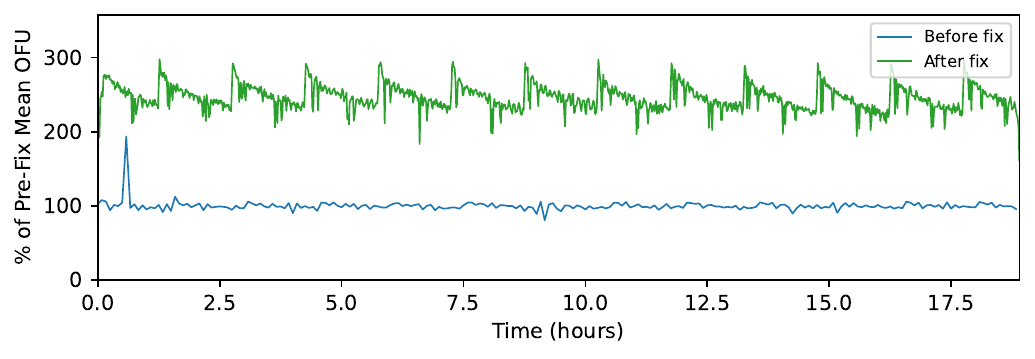}
  \caption{OFU before and after removing debug overhead, normalized to
  the pre-fix mean.}
  \label{fig:gear}
\end{figure}

Because OFU is model-architecture independent, it works for every training
experiment without requiring researchers to manually derive FLOPs per token.
This is especially valuable for labs working with novel architectures where
application-level MFU is either not calculated or calculated incorrectly, as
demonstrated in the internal case studies (Section~\ref{sec:case_studies}).
Even for smaller experiments where teams do not typically invest in
performance tuning, OFU provides zero-effort visibility that can surface
low-hanging-fruit misconfigurations.
A robot foundation model experimental training run on 256 H100 GPUs (32 nodes) observed lower OFU (Fig.~\ref{fig:gear}) than expected. Because OFU flagged the issue immediately, the team collected a
PyTorch Profiler trace to investigate.

The trace revealed that the environment variable
\texttt{TORCH\_DISTRIBUTED\_DEBUG=DETAIL} had been set and merged to the
main repository, which causes
PyTorch to inject \texttt{gloo:all\_gather} validation calls during every
NCCL collective operation. The before-fix trace contained 15,360
\texttt{gloo:send} events, 7,560 \texttt{gloo:recv} events, and 240
\texttt{gloo:all\_gather} events, all running over CPU-based Gloo
transport. These validation collectives serialized with the NCCL all-reduce
operations, dominating wall-clock time and leaving the Tensor Cores idle
for the vast majority of each training step.

After removing the debug flag, the Gloo
operations disappeared entirely from the trace and OFU improved by $2.5\times$
(Fig.~\ref{fig:gear}).

This case study illustrates two points. First,
it is often impractical for researchers to derive FLOPs per token for every
experiment, making issues like this difficult to surface without a
hardware-level metric like OFU.
Second, OFU is best understood as a coarse utilization signal. It identifies
\emph{that} a problem exists and quantifies its severity, but diagnosing
\emph{why} utilization is low requires profiling the job to identify the
specific bottleneck.

\subsection{Large-Scale Mixed-Precision Pretraining}
\label{sec:pretraining}

Extreme-scale training provided another opportunity to evaluate both the usefulness and practical
performance of OFU on real-world workloads. We incorporated OFU into a \toolRGS that monitors large-scale training jobs and takes corrective action to improve
productivity and reliability. Using this service, we monitored training runs for an internal research lab focused on scaling large language models and GPU-intensive deep learning workloads, pretraining a mixed-precision Mixture-of-Experts hybrid
Mamba-Transformer on 6{,}144-GPU Slurm jobs.

\begin{figure}[!t]
  \centering
  \includegraphics[width=\linewidth]{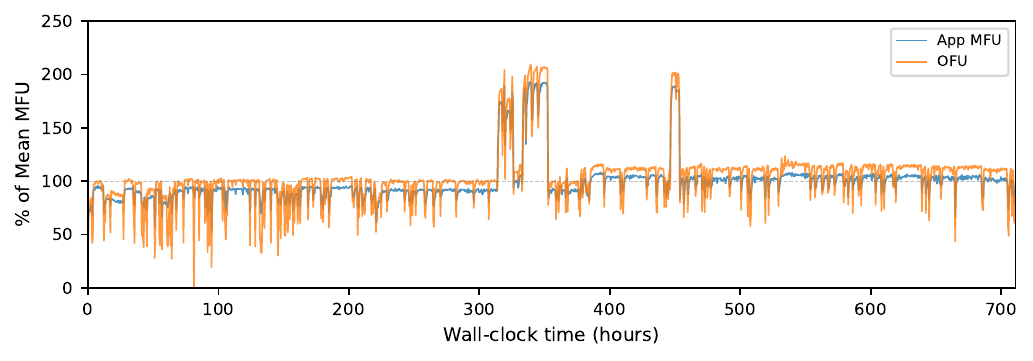}
  \caption{OFU and MFU relative to mean MFU on a pretraining at 6{,}144 GB200 GPUs.}
  \label{fig:6144_training_run}
\end{figure}

We evaluated 711 hours (${\sim}$30 days) of wall-clock training time on 6,144 GPUs (Fig.~\ref{fig:6144_training_run}). These
represent all jobs from the training run that used 6{,}144 GPUs and had both OFU data and
application-level throughput data emitted at least every 90 seconds. Because this workload mixes
multiple precisions (BF16, FP8, NVFP4), which have different hardware peak throughputs, the
single-precision denominator in~\eqref{eq:mfu} does not apply. Instead, we define an effective peak as the FLOPs-weighted harmonic mean of per-precision peaks:
\begin{equation}
    P_{\mathrm{eff}} = \frac{\sum_{i} F_i}{\sum_{i} \frac{F_i}{P_i}},
    \label{eq:peff}
\end{equation}
where $F_i$ is the FLOPs executed at precision $i$ and $P_i$ is the corresponding hardware peak
from~\cite{nvidia_gb200_product_page}. Application MFU is then computed as in~\eqref{eq:mfu} with
$P_{\mathrm{eff}}$ replacing the single-precision peak.

Across the 711-hour span, the point-by-point correlation between OFU and application MFU was $r = 0.718$, with short-term noise from clock sampling and transient workload phases reducing the agreement at individual time steps. When we average both metrics per job, the noise cancels out and the correlation across the 174 jobs rises to $r = 0.977$, indicating that OFU reliably distinguishes efficient jobs from inefficient ones.
OFU is precision-agnostic by construction.

The tensor pipe activity counter measures cycles
executing tensor instructions regardless of numeric format. This run provided a natural
test of this property, and provides empirical real-world backing for the results in~\ref{sec:gemm_validation}. Multiple debugging periods required switching from mixed precision (NVFP4, FP8, and BF16) to BF16-only. Observed compute throughput (TFLOP/s/GPU) remained roughly constant
between these modes. However, because BF16 has a lower theoretical peak than FP8 or NVFP4, the
effective peak for BF16-only periods was lower. With roughly constant throughput and a lower
denominator, application-reported MFU increased accordingly.
Figure~\ref{fig:6144_training_run} shows that OFU exhibited a corresponding increase
(both metrics are shown as a percentage of mean MFU in the figure).
In both mixed-precision and BF16-only modes, the two metrics agreed within 1 absolute percentage points on average, confirming that OFU correctly reflects precision-dependent utilization changes despite having no knowledge of the numeric format in use.

\subsection{World Foundation Model Training}
\label{sec:wfm}

An 8B-parameter world foundation model for generating physics-aware images and videos was trained on 256 GB200 GPUs. The run reported application-level MFU of 26\%, while OFU measured 34\%---a larger discrepancy than the 1--2 percentage points expected from tile-quantization overhead alone. Investigation revealed that the framework's FLOPs formula did not account for the additional forward-pass recomputation introduced by activation checkpointing~\cite{chen2016training}. With full activation checkpointing enabled, each training step performs roughly $4F$ FLOPs ($F$ forward $+$ $F$ recomputed forward $+$ $2F$ backward) rather than the standard $3F$, a 33\% increase. After correcting the FLOPs formula, application-reported MFU rose from 26\% to 33\%, aligning with the 34\% OFU to within 1 percentage point.

\section{Related Work}
\label{sec:related_work}

GPU utilization measurement spans three broad categories, each with
distinct trade-offs between fidelity, coverage, and deployment cost.

Performance profiling tools such as NVIDIA Nsight
Compute, Nsight Systems, PyTorch Profiler~\cite{pytorch_profiler},
DeepSpeed Flops Profiler~\cite{deepspeed_flops_profiler}, and
HPCToolkit~\cite{hpctoolkit} provide detailed per-kernel metrics
including instruction counts, memory bandwidth, and occupancy. These
tools offer the highest measurement fidelity but require per-job
instrumentation, impose runtime overhead, and are designed for targeted
profiling rather than continuous fleet-wide monitoring.

Framework-level MFU estimation, popularized by PaLM~\cite{chowdhery2022palm},
derives throughput from model-architecture FLOPs counts divided by
wall-clock time. Megatron-LM~\cite{nvidia2025megatronlm},
NeMo~\cite{nvidia_nemo}, and PyTorch Lightning~\cite{pytorch_lightning}
each implement variants of this approach. Google has published TPU~v4
MFU benchmarks using a similar methodology~\cite{google2023tpuv4mfu}.
MegaScale~\cite{jiang2024megascale} reports MFU at 10,000+ GPU scale,
and MLPerf~\cite{mattson2020mlperf} provides standardized training
throughput benchmarks across hardware platforms.
While these estimates require no hardware-level access, they depend on
manually derived FLOPs formulas that must be updated for each new
architecture---mixture-of-experts, latent-space routing, multimodal
pipelines---and can silently become incorrect as models evolve, as
demonstrated in our production case studies.

Hardware performance counters exposed through NVIDIA
DCGM~\cite{nvidia_dcgm_exporter} provide non-intrusive utilization
signals at negligible overhead. Prior work has used these counters
primarily for coarse utilization
metrics such as SM Activity~\cite{wu2021datacenter} rather than
as a first-principles MFU estimator. To our knowledge, OFU is the first
systematic study that derives, characterizes, and validates a
hardware-counter-based MFU proxy across multiple GPU generations and
precisions, with bounded accuracy established through controlled
experiments.

The roofline model~\cite{williams2009roofline} bounds throughput by
arithmetic intensity but requires per-kernel analysis and does not yield
a continuous fleet-wide utilization metric.

\section{Conclusion}
\label{sec:conclusion}

Fleet-wide GPU utilization measurement demands a metric that is
non-intrusive, precision-agnostic, and accurate across GPU generations.
We derived Overall FLOP Utilization (OFU) from first principles of GPU
floating-point execution pipelines, grounding it in Tensor Pipe Activity
and SM clock frequency---two signals universally available through
hardware performance counters---and characterized five sources of
divergence from application-level MFU: tile quantization and cuBLAS kernel selection,
floating-point precision portability, SM clock sampling noise,
Tensor Core clock domains, and non-tensor undercounting. Evaluation on 3{,}500 controlled GEMM
experiments across FP16, TF32, FP8, and NVFP4 on both H100 and GB200
demonstrated $\leq$2 percentage-point accuracy after tile-quantization
correction, and against 608 production training jobs OFU achieved
$r = 0.78$ correlation with application-level MFU while surfacing two
distinct framework-level FLOPs miscalculations. Deployed across large-scale
GPU fleets, OFU has detected a $2.5\times$ efficiency regression in
embodied-agent training, and when integrated into fleet-wide resilience and goodput
services, has surfaced performance drops invisible to
application-reported throughput. Our operational experience yields three lessons:
(1)~OFU is less error-prone than application-level MFU and can pinpoint
workloads with significant optimization opportunities;
(2)~OFU works well at very large scale, faithfully tracking mixed-precision
and lower-precision training---a typical optimization evolution---without
any code instrumentation;
(3)~instant, fleet-wide visibility enables rapid diagnosis of performance
bugs, leading to improvements as large as $2.5\times$ in our case studies.
Overall, our evaluation and operational experience indicate that OFU
can significantly lower the barrier to unearthing GPU inefficiency for AI workloads
on HPC systems.

\section*{Acknowledgments}
\noindent\textbf{AI-Generated Content Disclosure:}
AI writing tools were used for editorial assistance in preparing this
manuscript.

\bibliographystyle{IEEEtran}
\bibliography{paper}

\end{document}